\documentclass[conference]{IEEEtran}
\IEEEoverridecommandlockouts
\usepackage{cite}
\usepackage[numbers,sort&compress]{natbib}

\usepackage[utf8]{inputenc}
\usepackage{amsmath,amssymb}
\usepackage{graphicx}
\usepackage{bm}
\usepackage{enumerate}
\usepackage{comment}
\usepackage{xcolor}
\usepackage{url}
\usepackage{color,soul}
\usepackage{subcaption}
\usepackage{float}

\definecolor{light-gray}{gray}{0.8}

\usepackage{amsmath,amssymb,amsfonts}
\usepackage{algorithmic}
\usepackage[linesnumbered,ruled]{algorithm2e}
\usepackage{graphicx}
\usepackage{textcomp}
\usepackage{xcolor}
\usepackage{subcaption}

\def\BibTeX{{\rm B\kern-.05em{\sc i\kern-.025em b}\kern-.08em
    T\kern-.1667em\lower.7ex\hbox{E}\kern-.125emX}}

\makeatletter
\newcommand{\linebreakand}{%
  \end{@IEEEauthorhalign}
  \hfill\mbox{}\par
  \mbox{}\hfill\begin{@IEEEauthorhalign}
}
\makeatother

\begin{document}

\title{Cross-Cloud Data Privacy Protection: Optimizing Collaborative Mechanisms of AI Systems by Integrating Federated Learning and LLMs\\}

\author{

\small 

\begin{tabular}[t]{c@{\extracolsep{8em}}c} 

\textsuperscript{} Huaiying Luo & Cheng Ji\\
\textsuperscript{} College of Computing and Information Science & Siebel School of Computing and Data Science\\
\textsuperscript{}Cornell University & University of Illinois Urbana-Champaign\\
\textsuperscript{}New York, USA & Champaign, Illinois, USA \\
\textsuperscript{}hl2446@cornell.edu & chengji5@illinois.edu \\

\\

\end{tabular}
}

\maketitle

\begin{abstract}
In the age of cloud computing, data privacy protection has become a major challenge, especially when sharing sensitive data across cloud environments. However, how to optimize collaboration across cloud environments remains an unresolved problem. In this paper, we combine federated learning with large-scale language models to optimize the collaborative mechanism of AI systems. Based on the existing federated learning framework, we introduce a cross-cloud architecture in which federated learning works by aggregating model updates from decentralized nodes without exposing the original data. At the same time, combined with large-scale language models, its powerful context and semantic understanding capabilities are used to improve model training efficiency and decision-making ability. We've further innovated by introducing a secure communication layer to ensure the privacy and integrity of model updates and training data. The model enables continuous model adaptation and fine-tuning across different cloud environments while protecting sensitive data. Experimental results show that the proposed method is significantly better than the traditional federated learning model in terms of accuracy, convergence speed and data privacy protection.
\end{abstract}

\begin{IEEEkeywords}
Cross-cloud environment, Data privacy protection, Federated learning, Large-scale language models
\end{IEEEkeywords}

\section{Introduction}
In the field of cross-cloud data privacy protection, the combination of Federated Learning (FL) and Large Language Models (LLMs) has gradually become a hot topic. As data privacy and security concerns become more and more important, traditional centralized data storage and processing methods are increasingly being questioned. Traditional AI training methods often rely on centralizing all data stored on a single server or data center for processing \cite{seth2025}. 

This centralized data processing model is prone to the risk of data leakage and privacy violations, especially in the face of sensitive data, such as medical and health records, financial information and personal privacy data, and the centralized storage and sharing method can be attacked or misused during data processing \cite{zhang2024}. 

Therefore, how to carry out efficient AI training under the premise of ensuring data privacy and security has become the focus of current research. In order to solve this problem, federated learning emerged as an emerging distributed learning method, which effectively avoids the risk of data leakage and protects the privacy of data by training the model on the local device and sharing only model updates (such as weights, gradients, etc.) instead of the original data \cite{rasel2025}. 

The key benefits of federated learning are that it reduces data transfer through local computation, ensuring the privacy of each data source while still maintaining the accuracy and performance of the model. Especially in a distributed environment, federated learning can not only conduct data analysis without violating privacy protection regulations, but also has strong scalability and flexibility, and can be widely used in multi-party collaboration scenarios \cite{gafni2024, li2024advances}.

Notwithstanding the burgeoning advancements in large-scale language models (LLMs) across diverse domains \cite{ding2024enhance, deng2024composerx, ji-etal-2024-rag, Ji2025, yi2025score, jin2025adaptivefaulttolerancemechanisms, 10.1145/3627673.3679071, yang2025research}, particularly within the realms of natural language processing and generative tasks \cite{10628639, he2023t, he2024give, yang2024hades, liu2024mt2st}, a notable caveat exists: the training of these intricate models necessitates substantial computational resources and considerable data processing capabilities. Single-cloud platforms often face computing resource bottlenecks and latency issues, which cannot meet the high-performance requirements of LLMs training. In this context, cross-cloud federated training has become a new solution \cite{ramaswamy2024}. 

Cross-cloud federated training coordinates the computing resources of multiple cloud platforms and makes full use of the advantages of different cloud environments to complete large-scale model training tasks collaboratively. This cross-cloud architecture can distribute data and computing resources across multiple cloud platforms, effectively reducing the computing pressure on a single platform while improving resource utilization and training efficiency.

The advantageous resources of different cloud platforms can be fully integrated, and the latency and bottlenecks caused by insufficient resources on a single platform can be reduced \cite{jin2025scalability}, thereby accelerating the process of model training. In addition, cross-cloud federated training not only reduces the cost of training by sharing computing load among multiple platforms, but also avoids excessive centralized processing in the data set, further improving the level of data privacy protection \cite{guo2023}. Compared with a single cloud platform, cross-cloud training can more flexibly adjust resource allocation, further improve training efficiency, and reduce overall costs while ensuring data security.

With cross-cloud training, data privacy protection and efficient computing are guaranteed, especially when dealing with large-scale models, which can effectively solve the problem of computing bottlenecks and latency. The system using cross-cloud architecture can not only improve the collaboration ability of large-scale language models (LLMs) between different cloud platforms.

However, the training process of AI systems \cite{bosch2021engineering, li2024deception, li2024exploring, ding2024confidence, wang2024fine, yang2025data}, promotes the research and application of cross-cloud data privacy protection technology based on federated learning. With the continuous improvement of cloud computing infrastructure and the maturity of federated learning frameworks, cross-cloud federated training is expected to become one of the mainstream methods for AI model training.

\section{Related Work}
Lin et al. \cite{lin2024} point out that although LLMs excel in tasks such as code understanding, high-quality code data often has commercial or sensitive value, limiting its availability in open source AI projects. To solve this problem, the authors propose a governance framework with federated learning as the core, which aims to promote the joint development and maintenance of open-source AI code models while ensuring data privacy and security. 

Lazaros et al. \cite{lazaros2024} emphasize that federated learning not only enables multi-party collaboration while preserving privacy, but also transforms Internet of Things (IoT) systems into more collaborative, privacy-preserving and flexible frameworks. The study provides insight into understanding the role of federated learning in collaborative intelligence.

Yao et al. \cite{yao2024} discussed the challenges of fine-tuning and cued learning in federated settings, analyzed the challenges of model convergence and high communication costs caused by data heterogeneity, and proposed potential directions for future research. 

Tao et al. \cite{tao2025} proposed the FLFT approach, which aims to address the challenges of fine-tuning large-scale pre-trained language models in a distributed environment. FLFT incorporates federated learning (FL) strategies that allow multiple participants to share model update information while protecting data privacy, enabling collaborative fine-tuning.

Mawela et al. \cite{mawela2024} proposed a web-based federated learning (FL) automation solution that aims to simplify the deployment and management of FL tasks. The system provides a user-friendly web interface with support for FedAvg algorithms, allowing users to configure parameters for FL tasks through an intuitive interface, reducing reliance on programming and network architecture. 

Vadisetty et al. \cite{vadisetty2024} proposed an AI-generated privacy protection protocol for cross-cloud data sharing and collaboration. The core innovation of this research lies in the design of a set of protocol frameworks that combine federated learning, differential privacy, dynamic encryption, and context-aware policies, aiming to improve the privacy and security of data collaboration in a multi-cloud environment.

Yang et al. \cite{yang2025} proposed a novel cross-cloud data privacy protection framework that aims to address data privacy and security issues when training large-scale language models (LLMs) in multi-cloud environments. The framework uses a federated learning (FL) approach that combines homomorphic encryption, dynamic model aggregation techniques, and cross-cloud data orchestration solutions to enhance security, efficiency, and scalability.

\section{METHODOLOGIES}
\subsection{Federated Learning and Large-scale language models}
First, let's review the traditional federated learning architecture. In standard federated learning, multiple edge nodes, such as compute nodes in different cloud environments, train the model locally and then aggregate their updated information (rather than raw data) to a central server. Set the local loss function for each node to Equation 1:
\begin{equation}
\mathcal{L}_i(w_i) = \frac{1}{N_i} \sum_{n=1}^{N_i} \ell \left( f(x_{i,n}; w_i), y_{i,n} \right), \tag{1}
\end{equation}
where $w_i$ is the model parameter of the $i$-th node, $x_{i,n}$ and $y_{i,n}$ are the features and labels of the $n$-th sample on node $i$, respectively, and $f(\cdot)$ is the model prediction function, $\ell(\cdot)$ is the loss function, and $N_i$ is the number of local samples of node $i$.

The goal of federated learning is to optimize the global model by aggregating local updates, and the global update rule is Equation~\eqref{eq:fedavg}:
\begin{equation}
w^{(t+1)} = \sum_{i=1}^{K} \frac{N_i}{N} w_i^{(t)}, \tag{2} \label{eq:fedavg}
\end{equation}
where $w_i^{(t)}$ is the model parameter of the $i$-th node after the $t$-th round of training, $N$ is the total number of samples of all nodes, and $K$ is the number of nodes participating in the training.

The formula aggregates the model updates of each node through the weighted average, so as to obtain the optimization of the global model. In this process, the original data is not directly transmitted or exposed, ensuring the privacy of the data.

In the existing federated learning framework, we combine large-scale language models (LLMs) to enhance the training efficiency and decision-making ability of the model. 

Suppose we have a pre-trained model based on LLMs that is capable of extracting contextual and semantic feature information. The way we introduce LLMs in federated learning is by utilizing LLMs for feature augmentation when training a local model on each node. Set the output of LLMs to Equation 3:

\begin{equation}
z_{i,n} = \textit{LLM}(x_{i,n}),
\label{eq:llm_output}
\end{equation}
where $z_{i,n}$ is the context feature of the LLM output. These features are then fed into the local model of node $i$ for training, which in turn updates the local parameter $w_i$. Specifically, the local training objective function is Equation~\eqref{eq:local_llm}:
\begin{equation}
\mathcal{L}_i(w_i) = \frac{1}{N_i} \sum_{n=1}^{N_i} \ell \left( f(z_{i,n}; w_i), y_{i,n} \right). \tag{4}
\label{eq:local_llm}
\end{equation}

By introducing LLMs, we are not only able to leverage the local data of each node for model training, but also improve the model's ability to understand the context and semantics of the data, thereby enhancing the performance of the global model. The addition of LLM enables the model to better capture the deep relationships in the data when dealing with complex tasks.

\subsection{Secure communication layer}

To further enhance privacy, we introduce a secure communication layer into the model that encrypts model updates and training data to ensure that sensitive information is not exposed during federated learning.

Specifically, we encrypt the model update information of each node through Homomorphic Encryption. Suppose the cryptographic update information of node $i$ is Equation~\eqref{eq:encrypt}:
\begin{equation}
\widehat{w}_i = \text{Enc}(w_i), \tag{5}
\label{eq:encrypt}
\end{equation}
where $\text{Enc}(\cdot)$ indicates an encryption operation.

The encrypted model update information will be aggregated on the server side, and the global model will be updated to Equation~\eqref{eq:decrypt} through the decryption operation:
\begin{equation}
w^{(t+1)} = \text{Dec} \left( \sum_{i=1}^{K} \frac{N_i}{N} \widehat{w}_i \right). \tag{6}
\label{eq:decrypt}
\end{equation}

By introducing homomorphic encryption, we ensure that the data itself is encrypted even during the transmission of model updates, preventing malicious nodes or attackers from accessing the model parameters, thus effectively protecting the user's data privacy. 

Finally, we adapted and fine-tuned the model across clouds. Whenever a model is migrated from one cloud node to another, we retrain the model through an adaptation layer to adapt to the data characteristics in the new environment. Set the migrated model to Equation~\eqref{eq:finetune}:
\begin{equation}
w' = w + \Delta w, \tag{7}
\label{eq:finetune}
\end{equation}
where $w'$ is the fine-tuned model parameter, and $\Delta w$ is the parameter update obtained during the fine-tuning process.

By fine-tuning across clouds, we are able to ensure that the migration of models between different cloud platforms does not result in performance degradation, while maintaining the versatility and adaptability of the models.

\section{EXPERIMENTS}
\subsection{Experimental setup}
The experiment uses the Google Cloud BigQuery dataset, which contains diverse data from real-world business scenarios, including financial, medical, social media, and geographic information, which has high practical application value. The dataset includes structured and semi-structured data that mimics multiple data formats in the real world, and some of the data involves sensitive information, such as personally identifiable information and transaction records, making it suitable for studying how to protect data privacy across cloud environments.

We will use the following four advanced comparison methods including:

\begin{itemize}
    \item FedAvg (Federated Averaging): FedAvg is a classic federated learning method that aggregates the updated average of the local model into a global model. It is one of the most commonly used benchmarking methods in federated learning. 
    \item DP-FL (Differential Privacy Federated Learning): DP-FL combines differential privacy technology with federated learning to add noise to the transmitted model updates during model training, thereby further improving data privacy protection capabilities.  
    \item SMC-FL (Secure Multi-Party Computation Federated Learning): SMC-FL combines federated learning and secure multi-party computation (SMC) technology to ensure the security of model updates through cryptographic computing without directly exposing data to participants in the collaborative learning process. 
    \item HE-FL (Homomorphic Encryption Federated Learning): HE-FL uses homomorphic encryption technology to ensure that data is computed and model trained in an encrypted state. This approach protects the privacy of the data, especially when dealing with highly sensitive data, such as medical or financial data.
\end{itemize}

\subsection{Experimental analysis}
Figure 1 illustrates the privacy protection effect of different approaches under different privacy budgets (epsilon). As can be seen in Figure 1, the privacy protection effect of all methods improves as epsilon increases. FedAvg has shown stable privacy protection effect, and gradually improves with the increase of epsilon. DP-FL performs well in terms of privacy protection, but it fluctuates slightly compared to FedAvg, indicating that there is a certain instability between privacy and accuracy in the differential privacy mechanism.

\begin{figure}[h!]
  \centering
    \includegraphics[width=0.9\linewidth, height=0.45\linewidth]{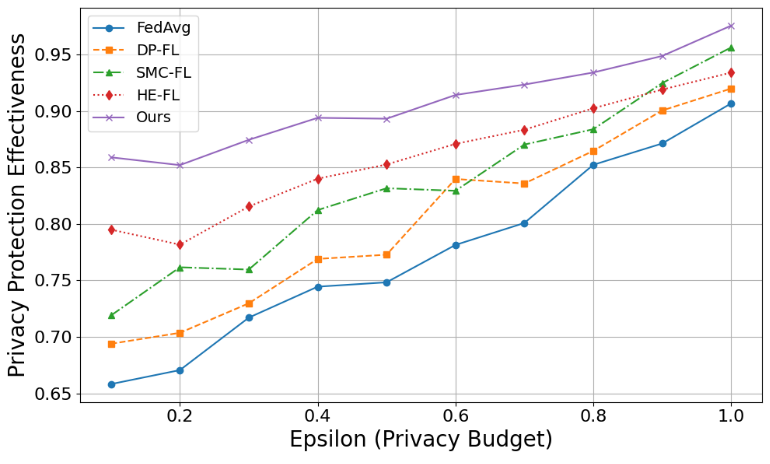}
    \caption{Differential Privacy Effectiveness Comparison for Different Methods}
  \label{fig:bar}
\end{figure}

The curve of SMC-FL is relatively smooth, indicating that secure multi-party computation provides a relatively balanced privacy protection effect. Although HE-FL provides a high privacy protection effect, it is highly volatile, which reflects that homomorphic encryption may affect the performance and stability of the model while improving privacy protection. Our $'$Ours$'$ method performed best of all methods, and its privacy protection improved rapidly with the addition of epsilon, showing that our method has a good balance between privacy protection and model accuracy.

As can be seen in Figure 2, the convergence speed (the number of iterations required to reach $85\%$ accuracy) generally decreases for all methods as the number of hidden cells increases. The $'$Ours$'$ method has the best convergence speed under all hidden unit configurations, and the number of iterations required is significantly less than that of other methods, indicating that it performs well in training efficiency. In contrast, FedAvg and HE-FL require more training cycles, especially in larger hidden unit configurations, showing slower convergence rates.

\begin{figure}[h!]
  \centering
    \includegraphics[width=0.9\linewidth, height=0.45\linewidth]{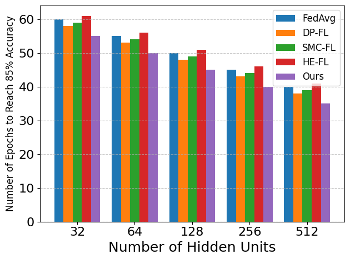}
    \caption{Convergence Speed Comparison for Different Methods}
  \label{fig:bar}
\end{figure}

As can be seen from Figure 3, as the learning rate gradually increases from $0.001$ to $0.05$, the training time of most methods decreases significantly, indicating that a higher learning rate helps to accelerate model convergence.

\begin{figure}[h!]
  \centering
    \includegraphics[width=0.9\linewidth, height=0.45\linewidth]{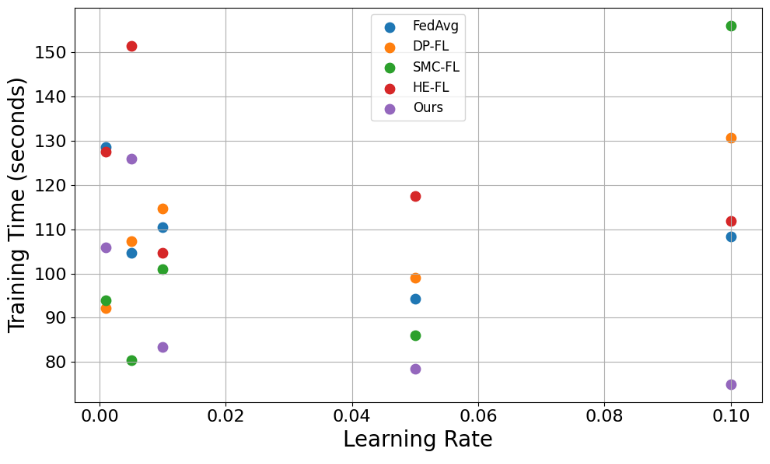}
    \caption{Training Time Comparison for Different Methods}
  \label{fig:bar}
\end{figure}

In contrast, FedAvg and DP-FL have a longer training time when the learning rate is low, but the convergence speed is significantly improved after a moderate increase in the learning rate. SMC-FL and HE-FL were more sensitive to changes in learning rate, and when the learning rate was too high ($0.1$), the training time increased, reflecting the instability of training.

\section{CONCLUSION}
In conclusion, by combining federated learning and large-scale language models, the cross-cloud collaborative training framework proposed in this study achieves the highest privacy protection effect in the differential privacy test, with the least iteration rounds, the fastest convergence, and the shortest and most stable training time under different learning rates, which is better than existing methods. As for the future, it can focus on dynamic privacy budget scheduling, adaptive communication compression strategies, and deep integration with trusted execution environment and hardware security module.

\renewcommand{\bibfont}{\footnotesize}

\footnotesize{
\bibliographystyle{IEEEtran}
\bibliography{main}
}

\end{document}